\documentclass[12pt]{article}
\usepackage{graphics}
\def\ni{\noindent}

\def\={\phantom{..} = \phantom{..}}

\def\+{\phantom{..} + \phantom{..}}
\def\>{\phantom{..} > \phantom{..}}
\def\<{\phantom{..} < \phantom{..}}
\def\-{\phantom{..} - \phantom{..}}

\def\lsb{\left[\,}
\def\rsb{\,\right]}
\def\lrb{\left(\,}
\def\rrb{\,\right)}
\def\lpar{\left\{\,}
\def\rpar{\,\right\}}

\def\ml{maximum likelihood}

\def\ro{R_0}
\def\vo{V_0}
\def\rl{R_l}
\def\rh{R_h}

\def\pe{p_e}

\def\no{\nonumber}
\def\be{\begin{equation}}
\def\ee{\end{equation}}

\def\bar{\begin{eqnarray}}
\def\ear{\end{eqnarray}}
\def\bad{\begin{equation} \begin{array}}
\def\ead{\end{array} \end{equation}}
\def\bmat{\left( \begin{array}}
\def\emat{\end{array} \right)}

\title{\bf \scalebox{1.5}{Stopping the SuperSpreader}\\ 
\scalebox{1.5}{Epidemic, Part III:}\\[1in] 
\scalebox{1.5}{Prediction}\\[2in]}

\author{W. David Wick$^1$\footnote{
Email:wdavid.wick@gmail.com}} 
 
\begin{document}
\maketitle
\pagebreak

\section*{Abstract}
In two previous papers, I introduced SuperSpreader (SS) epidemic models, offered some theoretical discussion of prevention issues, 
and fitted some models to data sets derived from published accounts of the ongoing MERS epidemic (concluding that a pandemic is likely). 
Continuing on this theme, here I discuss prediction: 
whether, in a disease outbreak driven by superspreader events, a rigorous decision point---meaning a declaration that a pandemic is imminent---can
be defined. 
I show that all sources of prediction bias contribute
to generating false negatives (i.e., discounting the chance of a pandemic when it is looming or has already started). 
Nevertheless, the statistical difficulties can be overcome by improved data gathering and use of known techniques that decrease bias.
One peculiarity of the SS epidemic is that the prediction can sometimes be made long before the actual pandemic onset, generating   
lead time to alert the medical community and the public.
Thus modeling is useful to overcome a false sense of security arising from the long ``kindling times'' characteristic of SS epidemics and certain political/psychological factors, 
as well as improve the public health response.    

\pagebreak

\section*{Text}

Again I avoid the standard journal format.
The only prerequisites for reading the main text of this paper are some familiarity with common statistical terms or phrases
such as ``bias'' and ``maximum-likelihood estimation,'' which I assume most readers will recall from
a college statistics course or will look up on Wikipedia.
I will relegate the statistical heavy-lifting to an appendix.

Perhaps the single most important contribution a mathematical modeler can make for public health would be to invent a tool that could predict the occurrence 
of a deadly world-wide pandemic of an infectious disease. As is well-known, most such diseases derive from zoonotic transmission events (meaning from an animal host in which the disease organism
is endemic, or is currently causing an epizootic, to humans); examples include plague, influenza, Ebola, HIV, SARS, and MERS. When a pathogen jumps species there is a likely requirement of certain 
genetic modifications before it can cause efficient transmission in the new host population. The implication is that prediction is primarily a business for geneticists---but, at present,
identifying the relevant mutations in the pathogen genome prior to their appearance seems beyond geneticists' capabilities. Perhaps that situation will soon change; in the mean time, it is up to
epidemiologists to use case data and contact tracing to declare the moment when the pandemic is inevitable, or has already begun. But is this even possible for the SuperSpreader (SS) epidemic?

\begin{figure}
%\resizebox{5in}{5in}{figure}
\rotatebox{0}{\resizebox{5in}{5in}{\includegraphics{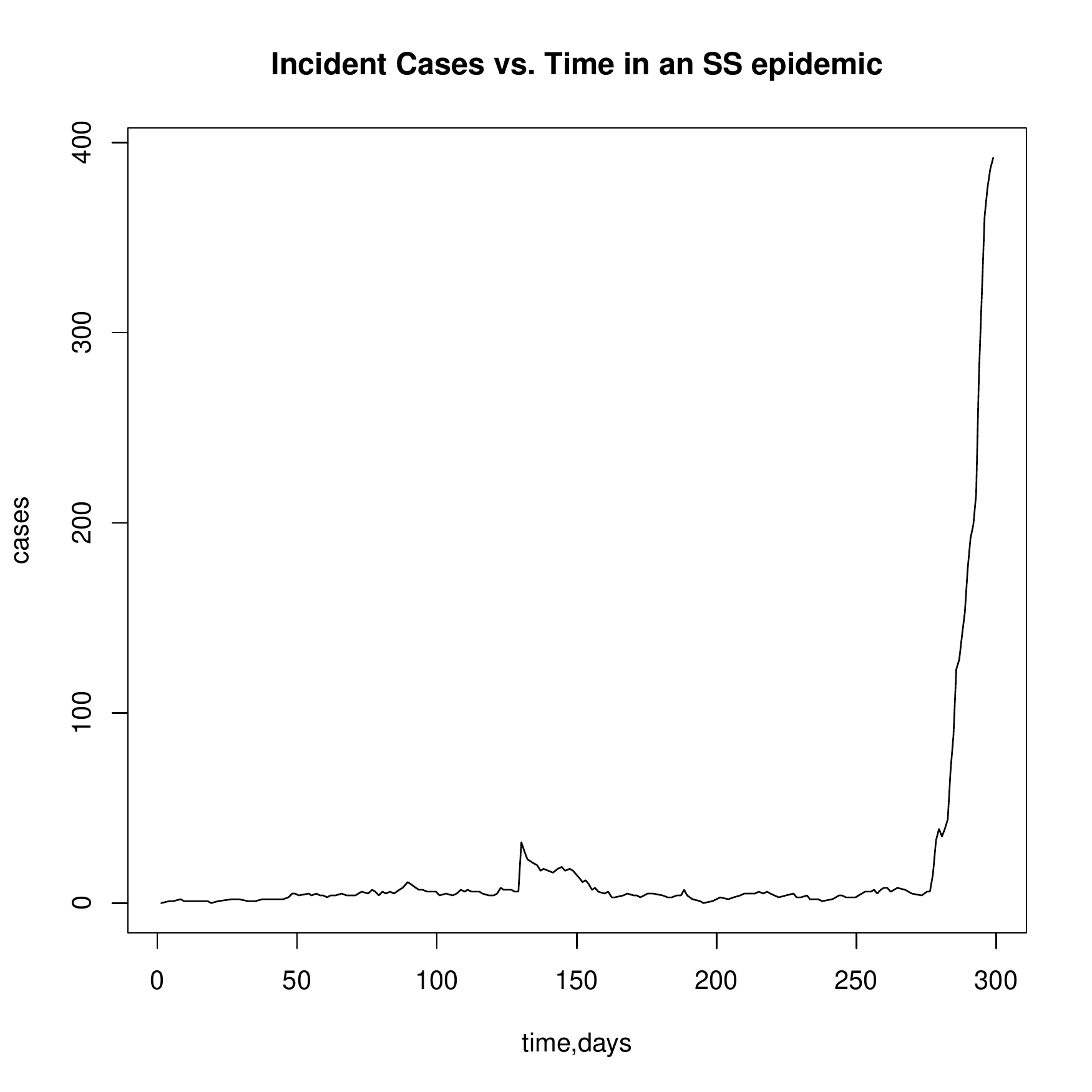}}}
\caption{Typical ``run'' of a simulated SS epidemic (parameters in text). The timing of events is highly variable in different runs.\label{fig0}}
\end{figure}

As I showed in previous papers, \cite{augustpaper},
\cite{maypaper}, an SS epidemic is unlike an influenza or measles epidemic, for which classical, deterministic mathematical models of the SIR (susceptible-infected-recovered) type are appropriate.
The SS epidemic has random outbursts and lulls, periodic re-introductions from the zoonotic source, and a long ``kindling'' period before the pandemic begins. 
(See Figure \ref{fig0} for a typical
simulated epidemic trajectory.)
Moreover, the case mortality for SARS and MERS is much higher than for influenza or measles (16 percent for SARS and 40 percent for MERS, as I write, as opposed to 2 percent for the 1918 flu), 
so by the time susceptibles drop limiting the pandemic,
the world would have suffered hundreds of thousands to millions of deaths. 
Thus the key issue is not predicting the size of the epidemic, but whether it is looming and we must quickly find a way to stop it.  

In paper II of the series, I used the information available to me (case-incidence data) to fit some SS models to MERS data through April 2014 and asserted that a pandemic was likely.
Since the data was less than optimal, I used a ``generic'' technique from pattern recognition to find models that fit. 
This work is more illustrative, so I will assume available detailed contact data of just the type we would wish to have in order to estimate model parameters, including $\ro$, the average
number of secondary (directly-infected) cases due to one primary case (which could be sporadic, from zoonotic transmission, or the secondary of yet another primary). 
In some instances this kind of who-infected-whom data will be readily available through screening, e.g., for health care workers exposed to a patient 
(who have comprised a large fraction of infections in the MERS epidemic), or passengers on a plane with the primary case; but for others it will be difficult, 
e.g., for community-derived infections. 
I'll indicate the (mock) ``data,'' which will be generated in simulations, by $\{n_1, n_2, n_3, \cdots n_N\}$, 
where $n_i$ stands for the number of secondary cases caused by the i-th infected  
case.\footnote{With 
complete contact data the details of infection
clusters could be worked out, as assumed in \cite{lancet} and \cite{ploscb}, but as the distribution of secondaries for each primary is the main issue, examining ``cluster sizes'' or 
``numbers of clusters'' implies an inefficient use of available information.} 
Of course, some contact data would be missing, resulting in misclassification of secondary cases as sporadic, 
causing undercounts of secondaries; I'll discuss the impact of this problem later.

For the model I choose a rather extreme two-level SS example\footnote{For mathematicians: the technical description 
is a continuous-time, multi-type branching process with two types, non-exponential waiting times, and 
non-Poisson offspring distributions. See paper I.} that appeared in the model fit of paper II; 
the parameters were: $\rl$ (low-level infectiousness) $= 0.3$, $\rh$ (high-level) $= 55$, and $\ro = 1.4$. 
In addition, the noninfectious and infectiousness periods were assumed lognormal with standard deviation of twice the means. 
Both the span of reproductive numbers and the large variance of the infectious period generate
``extra-Poisson variation'' (terminology from \cite{book08}; $\vo \neq \ro$, where $\vo$ is the variance of the secondaries) 
and strongly affect the statistical issues, as we will see. 

Since $\ro$ is the parameter that will drive the prediction, how would we estimate it from the secondaries data? Since we are presented with an i.i.d. sample, 
the obvious (non-parametric) approach is simply:
\def\rohat{\hat{R}_0}

\be
\rohat = {1\over N}\,\sum_{i=1}^N\, n_i.
\ee

\begin{figure}
%\resizebox{5in}{5in}{figure}
\rotatebox{0}{\resizebox{5in}{5in}{\includegraphics{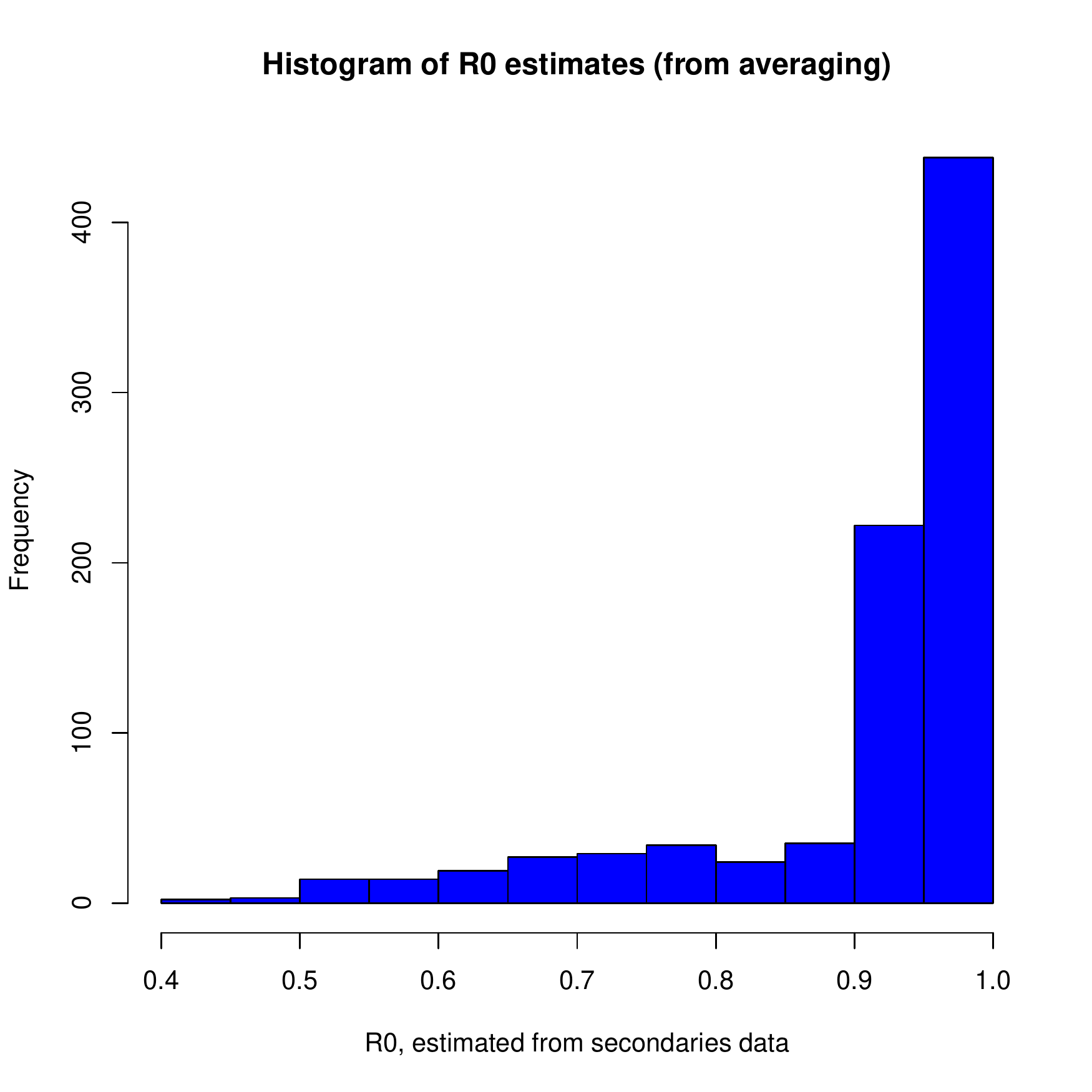}}}
\caption{$\ro$ estimated the easy way, from 1000 runs of the SS model with 2000 cases, showing downward bias.\label{fig1}}
\end{figure}

However this estimator is highly biased downward, as is seen in Figure \ref{fig1}. All the estimates were less than one (meaning no pandemic).

What aspects of this problem might contribute to generating this peculiar bias towards low $\ro$ estimates? First, data 
points that one would normally regard as outliers, and might even drop from the analysis,
namely the superspreader events, are in fact informative here. Second, there is skewing of distributions from the lognormal infectious periods. 
Finally, there is ``right censored data,'' which is a peculiarity of trying to estimate parameters during an ongoing epidemic. The simulations used for Figure \ref{fig1} 
were halted at 2000 total cases, on the logic that passing some threshold case load (hopefully smaller, but I will return to that issue below) 
is likely the moment when we would prepare the statistical analysis. However, SS epidemics are particularly prone to sudden explosions.
Thus many people may have been infected in the previous weeks---but then there would not have been time for them to generate as many secondary cases as would be expected
(and any investigative or reporting lag would exacerbate this problem). 

\begin{figure}
%\resizebox{5in}{5in}{figure}
\rotatebox{0}{\resizebox{5in}{5in}{\includegraphics{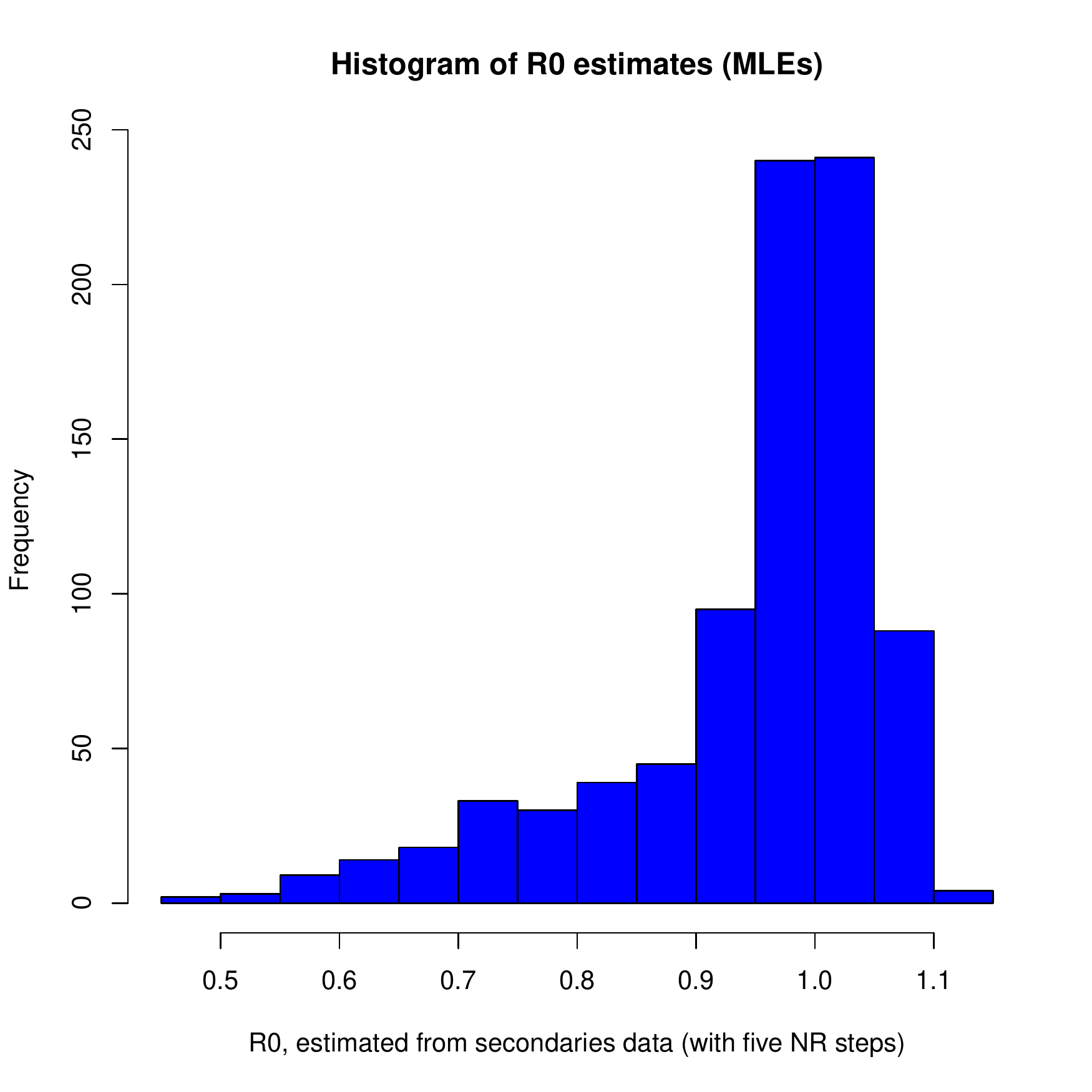}}}
\caption{$\ro$ estimated using maximum-likelihood model fits; same runs as in Fig. 1\label{fig2}}
\end{figure}

\begin{figure}
%\resizebox{5in}{5in}{figure}
\rotatebox{0}{\resizebox{5in}{5in}{\includegraphics{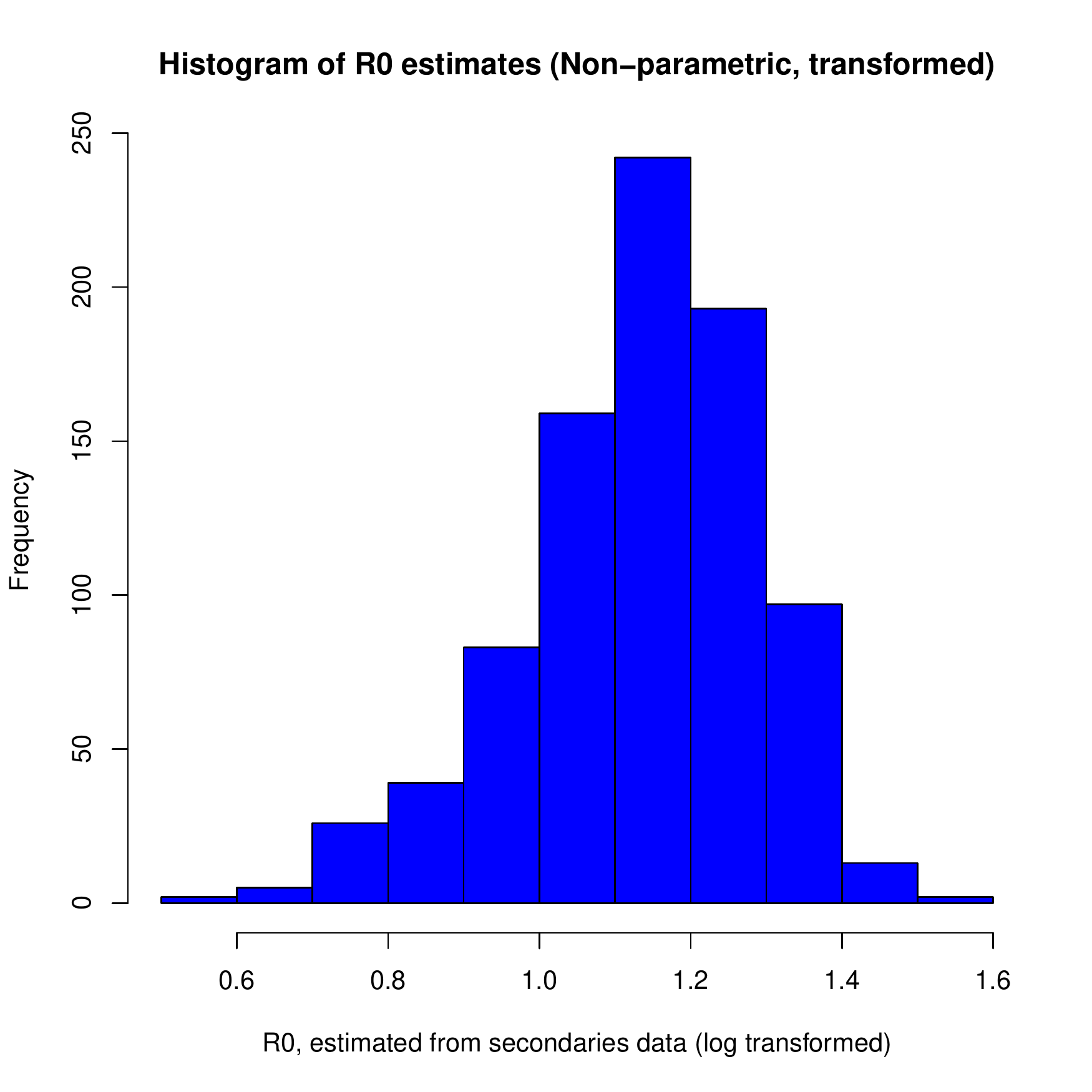}}}
\caption{$\ro$ estimated from non-parametric log-transformed secondaries data and model fitting; same runs as in Fig. 1\label{fig3}}
\end{figure}

What can we do about these problems? 
As a second try, I adopted the famous, asymptotically optimally-efficient, maximum likelihood method to fit the model. 
This made some improvement, see Figure \ref{fig2}.
Maximum likelihood is well known to be biased in small samples, with bias of $\hbox{O}(1/N)$, where $N$ here means the number of primary cases in the simulation (2000).
One might expect it to be small, but apparently something about the data still causes trouble.\footnote{Similar 
underestimation bias should afflict the Bayesian method of estimating $\ro$ described in \cite{ploscb}, 
as it is based on a likelihood. For more on this issue, see subsequent footnote.}
When faced with the problem of estimating skewed distributions, one is directed to try transforming the data. So I replaced $n_i$ by $\log(1+n_i)$ and fit the three-parameter model,
again assuming sufficient data, by equating model-averaged $\log(1+n)$ to empirical averages, see the Appendix for details. This worked much better; see Figure \ref{fig3}.

\begin{figure}
%\resizebox{5in}{5in}{figure}
\rotatebox{0}{\resizebox{5in}{5in}{\includegraphics{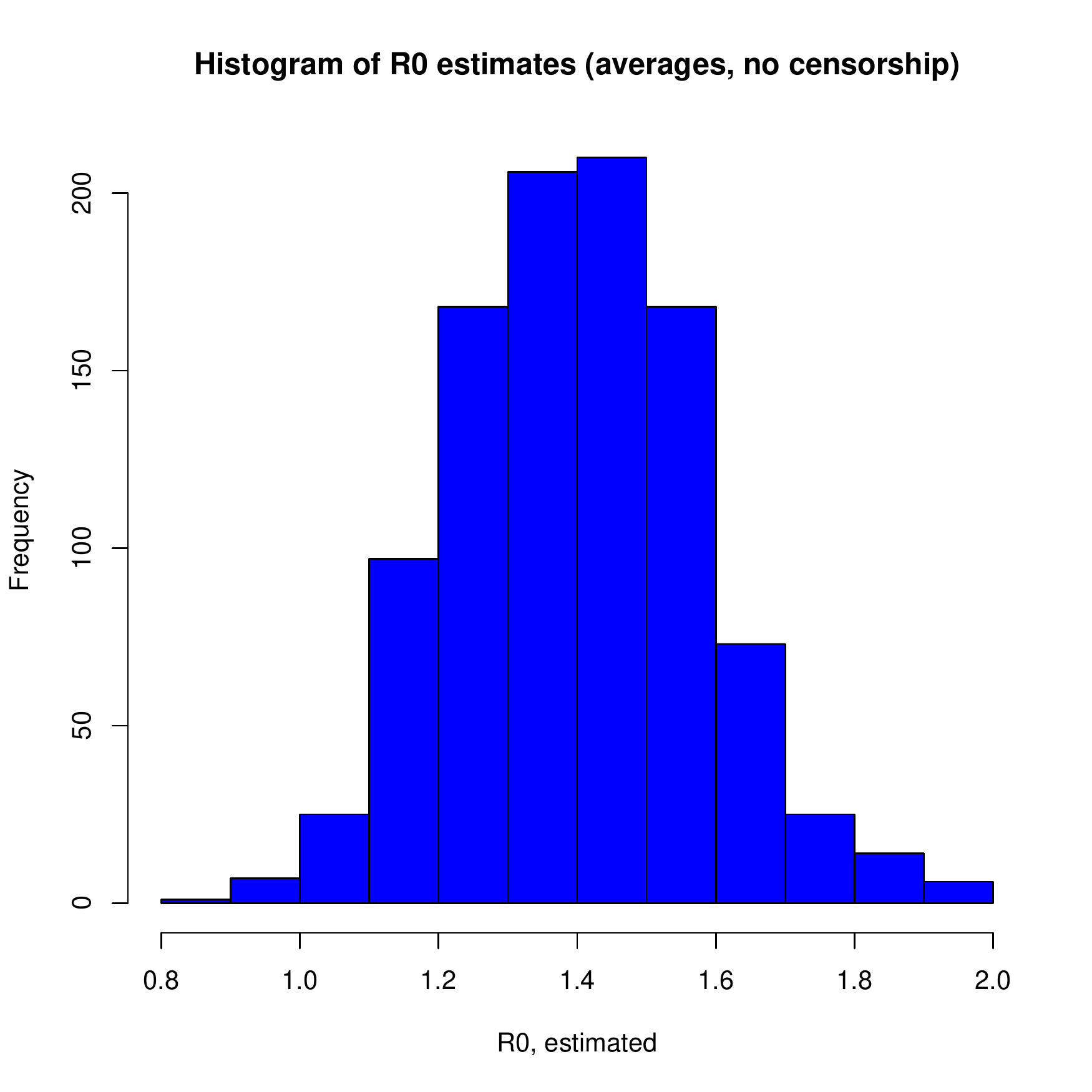}}}
\caption{$\ro$ estimated from averages of secondaries, without running the branching process (so no right censorship). 1000 runs, with 2000 primaries per run.\label{fignosimulation}}
\end{figure}

However, clearly bias remained despite using these statistical tricks.  
What about censorship? I simulated secondaries (for 2000 primaries) directly from the probability rule\footnote{See formulas (\ref{dist_eqns}) of the Appendix; 
thus the computer chose the infectiousness type, then a normal r.v., $Z$ and a corresponding lognormal infectious period, 
then ran a Poisson process to the end of that period, for 2000 cases, 1000 times.}  
with the same parameters as the model, but without running the branching process
and found Figure \ref{fignosimulation}.  Thus, the most important factor biasing the data was not skewing or outliers, but right-censorship. 
This censorship may occur in practise because: the primary is still in the asymptomatic (non-infectious) or infectious period, or some secondaries are still asymptomatic
and were not picked up through contact-tracing and screening (by e.g., RT-PCR), when the prediction was made. 
However, we pay a major cost for dropping cases, by 
lowering the power of the test.\footnote{For example, if I had dropped the last two weeks worth of cases from April 2014 in the model-fit of paper II, I would not have
been able to make the claim that $\ro > 1$ then.} Although there are well-known methods for improving estimation in the presence of censorship (by imputation of missing entries),
perhaps a preferable approach is to get better data! Now let us assume that, in addition to secondaries data (the $\{n_i\}$), we have available the times of onset and termination of infectiousness 
for the primary cases. This is not a great hardship; for SARS or MERS, the infectious period may coincide with the symptomatic period, 
and the endpoint of infectiousness with either the patient's death or
isolation. This kind of data could be obtained from patient records and so should be easier to derive than contact data. 

\begin{figure}
%\resizebox{5in}{5in}{figure}
\rotatebox{0}{\resizebox{5in}{5in}{\includegraphics{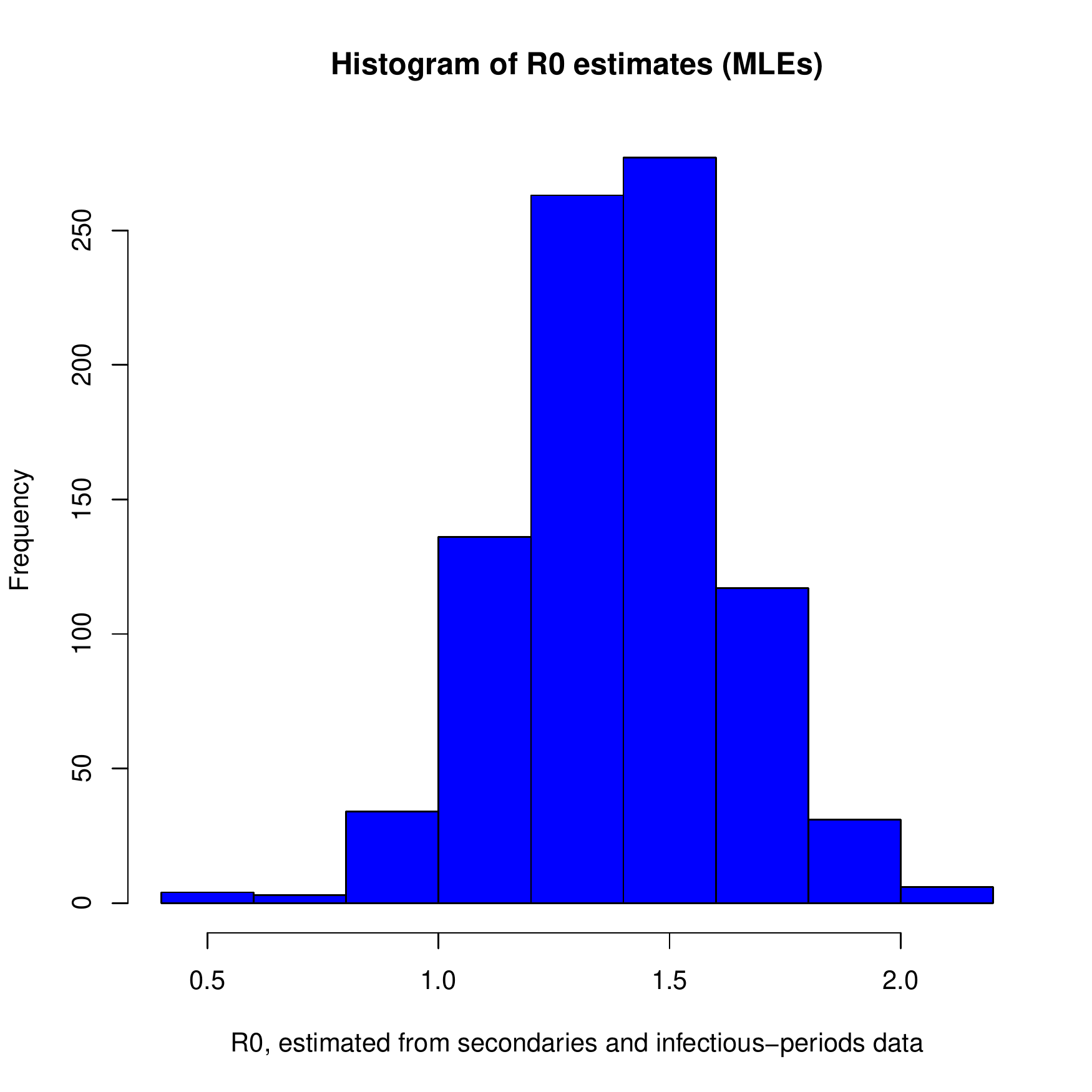}}}
\caption{$\ro$ estimated (MLEs) from secondaries and infectiousness-periods data (2000 cases).\label{figtimes1}}
\end{figure}

So I assumed (and stored in simulations) 
infectiousness periods; the data now consists of $\{n_i,\tau_i\}$, omitting cases where infectiousness onset was censored and if termination was, 
setting: $\tau_i =$ final time minus onset.  
(This choice automatically drops primaries still in the non-infectious period, while keeping secondaries infected before the cut-off time, even though they may take a few more weeks to locate.)
With this amplified data-set, maximum-likelihood analysis is appropriate (see Appendix). 
Moreover, we do not have to model the infectious period.\footnote{I adopted lognormal for the simulations and assumed it for fitting purposes, 
but no data sets available to me influenced that choice. It just seemed reasonable.}
See Figure \ref{figtimes1}.

Table 1 compares statistical properties of the four methods mentioned above. 
The log-transformed with secondaries data and the MLE with both secondaries and infectious-period data roughly tied on standard error, but the latter had much smaller bias.
\pagebreak 
\def\hb{\hfill\break}
\centerline{}
\centerline{}
\centerline{}
\centerline{Table 1. Statistical Properties of Four Estimators of $\ro$ for an SS epidemic.$^{1}$}
\begin{center}
\begin{tabular}{|c|c|c|c|c|c|}\hline
Method & Figure & No. runs$^{2}$ & Bias & Variance & SE$^{3}$ \\ \hline
averages (S)$^{4}$ & \ref{fig1} & 1000  & -.586  & .0369  & .617 \\ 
ML (S) & \ref{fig2}  & 852  & -.464  &  .0126 &  .478   \\ 
log-transformed (S) & \ref{fig3}  & 848  & -.264  & .0306 & .317   \\ 
ML (S\&P)$^{5}$ & \ref{figtimes1} & 848  & .0398  & .100  & .319 \\ \hline
\end{tabular}
\end{center}
$^{1}$ Parameters as in text; 1000 runs stopped at 2000 cases.\hb
$^{2}$ Some runs had too few secondary ``high-level'' ($n > 5$) cases, or NR did not converge.\hb
$^{3}$ Standard error; square root of: (Bias squared plus Variance).\hb
$^{4}$ Secondaries data only.\hb
$^{5}$ Secondaries plus infectiousness periods.\hb
\centerline{}
\centerline{}
\centerline{}

\begin{figure}
%\resizebox{5in}{5in}{figure}
\rotatebox{0}{\resizebox{5in}{5in}{\includegraphics{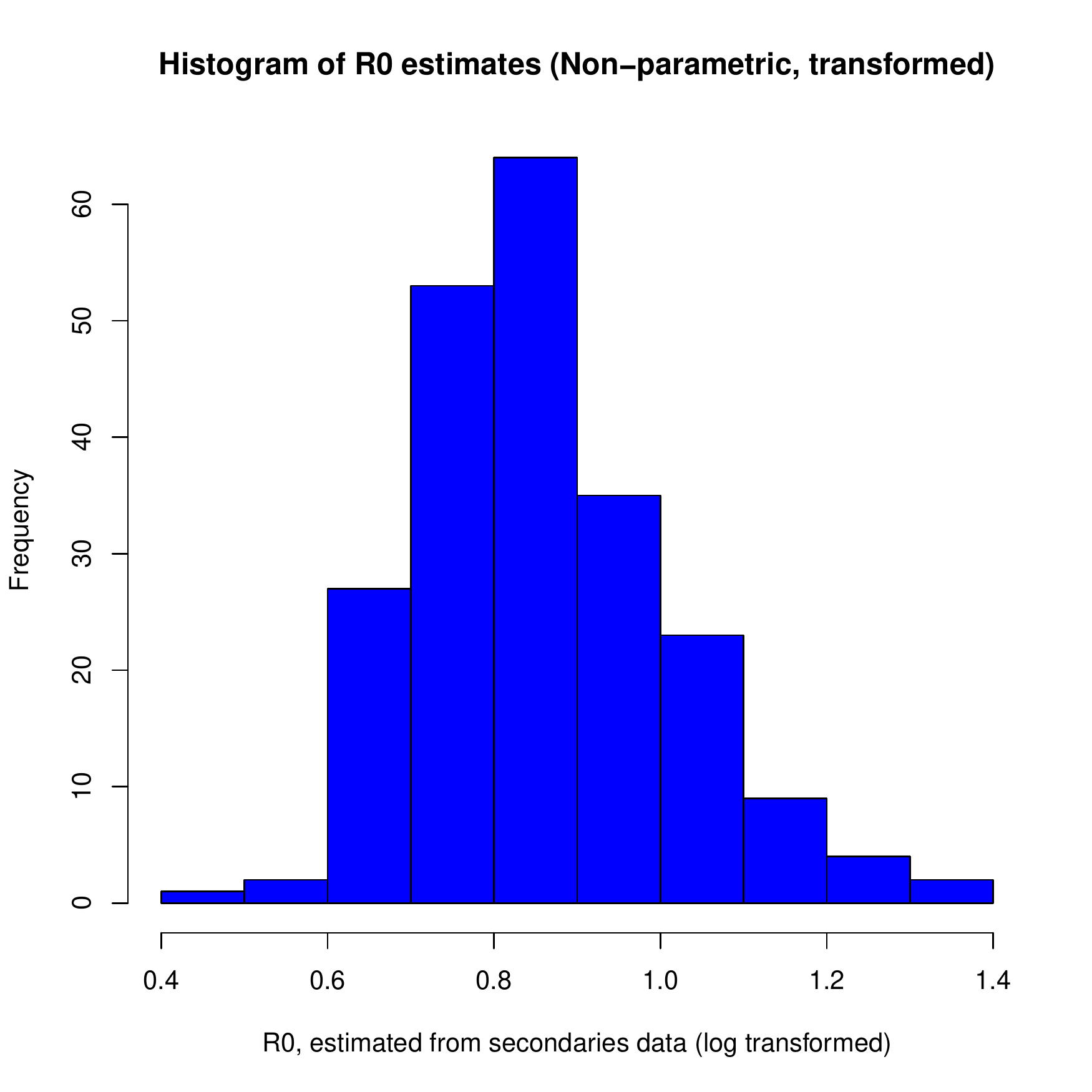}}}
\caption{$\ro$ estimated from log-transformed secondaries data and model fitting, but model $\ro = 0.8$. 1,000 runs, but all but 220 were dropped for lack of 
superspreader events, see text.\label{fig5}}
\end{figure}

\begin{figure}
%\resizebox{5in}{5in}{figure}
\rotatebox{0}{\resizebox{5in}{5in}{\includegraphics{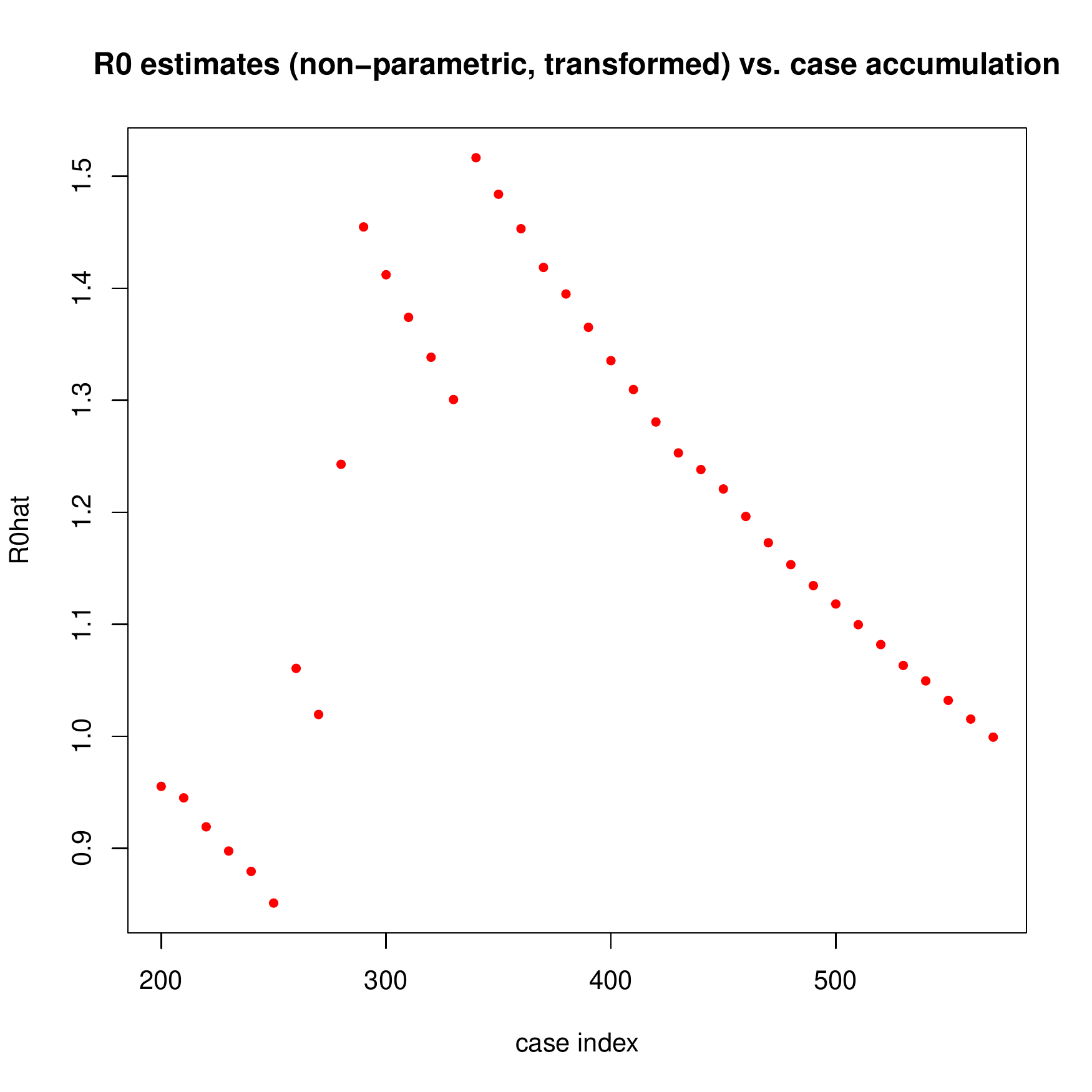}}}
\caption{$\ro$ estimated (log-transformed method) from data at early time points in the epidemic.\label{fig4}}
\end{figure}

What about estimation in an epidemic with $\ro < 1$? See Figure \ref{fig5}.
Although there is some overestimation of $\ro$, it is very minor. In fact, in only 220 of 1,000 runs were there enough high-level events (with more than five secondaries)
to even carry out parameter estimation; these runs are unlikely to be judged pandemics, leaving only about 40 out of 1,000 (or 4 percent) which would generate a false positive
(crying ``Pandemic!'' when there wasn't one).  
Note that all forms of statistical bias tend towards generating false negatives, i.e., a false sense of security that we will escape the horrors of a world-wide
epidemic.\footnote{The cluster-based methods of 
Blumberg and Lloyd-Smith, \cite{ploscb}, are not applicable to SS epidemics (as defined in my papers), 
despite the authors' assertion that they have incorporated ``superspreader'' events. As pointed out in an appendix to paper I, their negative-binomial
offspring distribution interpolates between a Poisson and a geometric; but the latter is just a Poisson process run over an exponential infectivity time. 
Neither endpoint is biologically appropriate for an SS epidemic. 
Moreover, they tested 
their methodology only for the case $\ro < 1$ (hence their ``stuttering chains''), 
in which the skewness, outlier, and censorship problems under discussion here do not arise. 
Using their technique to argue that $\ro < 1$, as in \cite{lancet}, may be circular reasoning.
Nevertheless, using my models and kNN fitting technique, in August 2013 I did agree with 
the Lancet authors that $\ro$ for MERS was less than one at that time. But, looking at Fig. \ref{fig4}, 
I now wonder whether we were all mistaken and that period was merely the ``kindling'' phase
of a pandemic.}

\begin{figure}
%\resizebox{5in}{5in}{figure}
\rotatebox{0}{\resizebox{5in}{5in}{\includegraphics{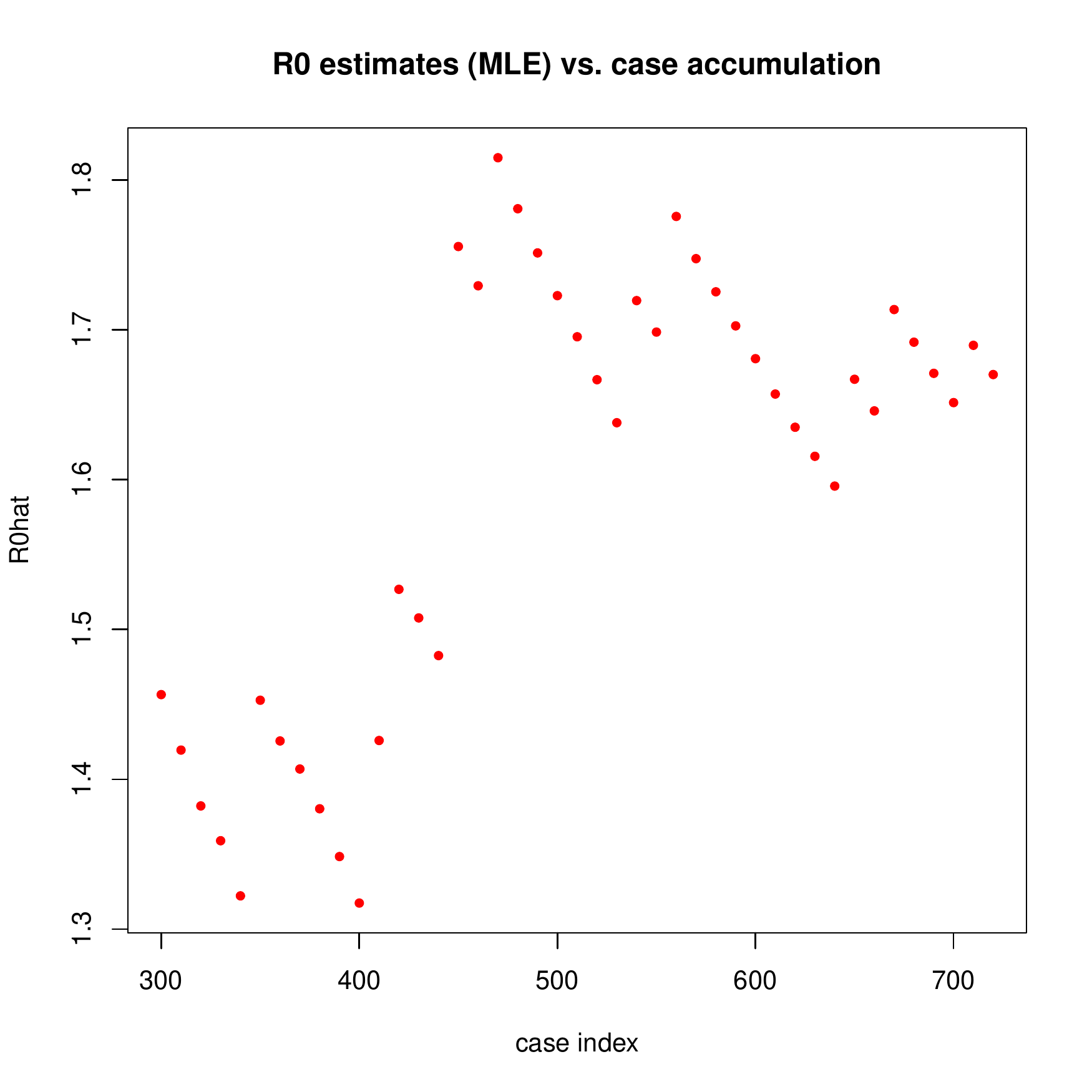}}}
\caption{$\ro$ estimated (MLEs) from secondaries-plus-infectious-periods data at early time points in the epidemic.\label{timesfig4}}
\end{figure}

The MLE method with secondaries and infectious-periods data did a good job at overcoming statistical bias. 
However, there is a fourth source of bias---going in the same direction---shown
in Figures \ref{fig4} and \ref{timesfig4}.
Figure \ref{fig4} shows estimations of $\ro$ derived from secondaries data by the log-transformed method, and Fig. \ref{timesfig4} from secondaries plus infectious periods by the MLE method, 
up to the number of cases shown on the x-axis, early in the epidemic. 
Note how the estimates rise abruptly and then fall; what is happening is that the estimates are driven upwards by an SS event and then fall as more non-SS events accumulate.
Note also how, ignoring the sawtooth pattern, the general trend is upwards. (Fig. \ref{timesfig4} looks better, but both figures have very few dots because of too few ``high-spreader events,''
e.g., with $n > 5$, to run the nonparametric estimators, early in the epidemic.) 
Call it ``temporal bias,'' by which I mean bias due to estimating too early; it points in the same 
direction as the statistical biases we saw in earlier diagrams, towards underestimating $\ro$.\footnote{Which raises an interesting questions about estimating with long-tailed distributions:
is there some general statement of the form, ``If your first estimate is that $\rohat$ = something, is it probably x percent higher?'' I hope to address this question in a later publication.} 

In addition to techniques for decreasing bias, we need some way of assessing confidence in our prediction. After some experimentation with popular methodology 
(e.g., resampling and the ``jackknife'') for
constructing confidence intervals\footnote{Fig. \ref{figtimes1} suggests using ordinary, normal-distribution based CIs, but recall that figure
assumed 2,000 cases; with fewer we can not be confident of the normal law.} which worked poorly at least in my implementation, 
I recurred to the ancient Fisherian ``reject the null hypothesis'' philosophy.
Fisher's method has the advantage that it addresses our major concern: that the declaration of a pandemic might be due merely to chance. 
However, where modeling is involved there is a pitfall in Fisherism:
one must not adopt absurd models (e.g., here, uniform or Poisson cases) as ``null models,'' because by that methodology any desired model can be ratified. 
I adopted for the null a model with the same observed frequencies of lower/higher types, then selected $\rl$ and $\rh$ at random but restricted so that $\ro = 1$.
By simulation we obtain a probability that the observed $\ro$ would be equaled or exceeded in non-pandemic SS models, i.e., a ``type-I error.'' (See the Appendix for details.)
That permits an announcement with confidence, e.g., ``The probability of a pandemic has now exceeded (say) 90 percent.'' 

As I showed in paper I of this series, SS epidemics can have a high probability that a given infection chain started from one ``index'' case goes extinct. For example, with the
model selected here for illustrative purposes, the extinction probability, $\pe$, was about 0.991.\footnote{Extinction probabilities 
are not available analytically except for unrealistic Markovian models, see 
paper I appendix; so they were computed by simulating 1,000 epidemics.} 
As a consequence of these high extinction rates,
there may be a long period in which sporadic cases appear but the infection-chains peter out, which again biases us towards optimism that we have escaped the worst.
Now consider prediction. Suppose that, as the epidemic progresses, we apply our statistical analysis to 
case/contact data and, at some point, obtain an estimated $\ro$ that is greater than one, with a p-value less than 
0.1\footnote{Not 0.05; the goal isn't to get a paper published, but to stop a pandemic!}
But when will the pandemic start? 

Here we must take into account Fig. \ref{fig4}, which suggests that our first indication of a pandemic will probably coincide with an SS event. So we should estimate the time-to-pandemic
by:

\be
\hbox{time-to-pandemic} \= p_e^n\,\,\left({\hbox{mean sporadic interval}\over 1 - p_e}\right).\label{time_eqn}
\ee

\ni where $n$ denotes the number of secondary cases of the case appearing at the instant in question (which will admittedly take some extra time to locate).
The first factor takes into account that, if one of the chains begun by these secondaries goes pandemic, 
the pandemic is already underway (so an omitted term: $(1 - p_e^n)\times 0$), 
and otherwise we wait until
some future sporadic case generates the pandemic. 

\begin{figure}
%\resizebox{5in}{5in}{figure}
\rotatebox{0}{\resizebox{5in}{5in}{\includegraphics{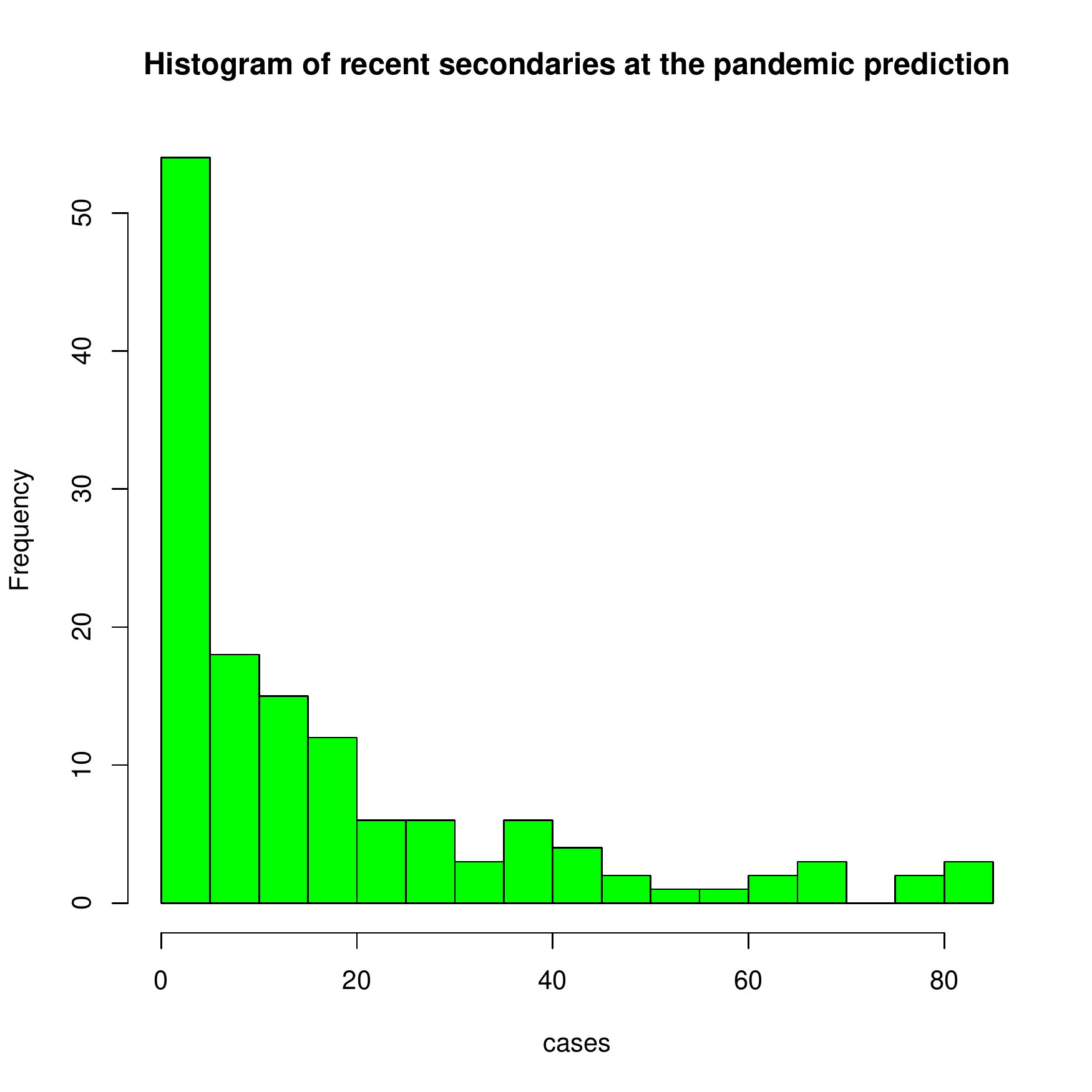}}}
\caption{``n'' at the pandemic prediction (MLE method). \label{fig6}}
\end{figure}

\begin{figure}
%\resizebox{5in}{5in}{figure}
\rotatebox{0}{\resizebox{5in}{5in}{\includegraphics{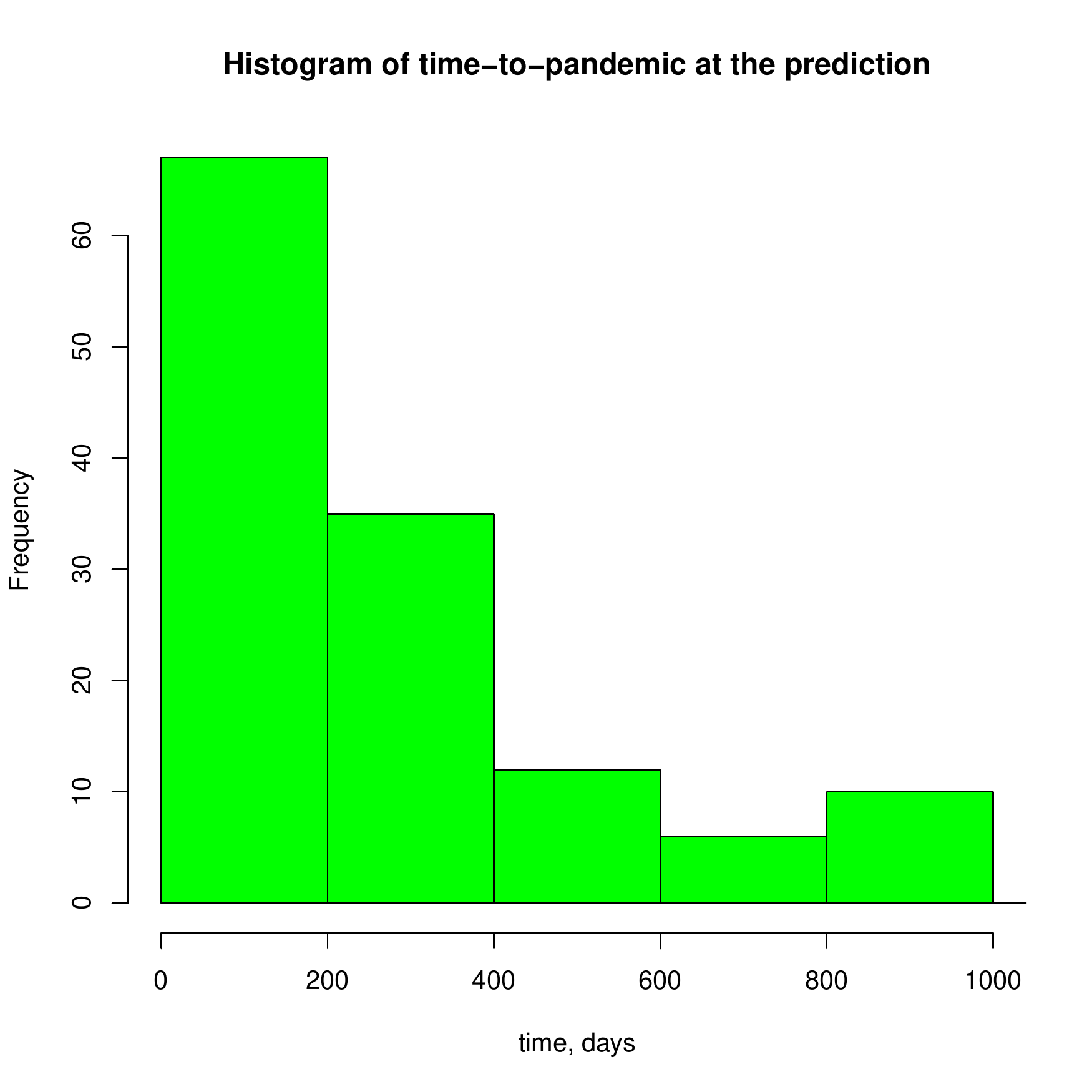}}}
\caption{``Time-to-pandemic'' at the prediction (MLE method, using last ``$n$''). \label{fig7}}
\end{figure}

\begin{figure}
%\resizebox{5in}{5in}{figure}
\rotatebox{0}{\resizebox{5in}{5in}{\includegraphics{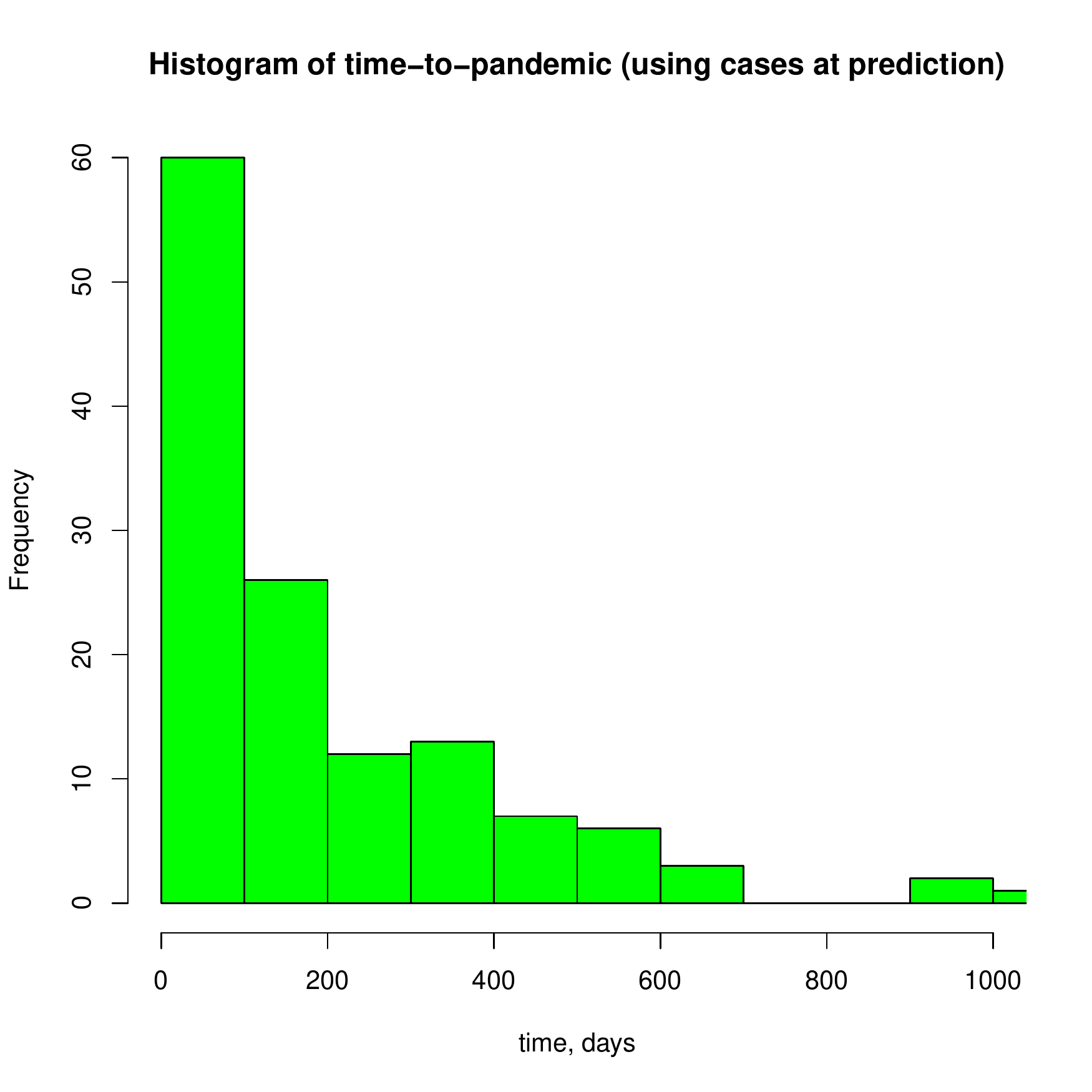}}}
\caption{``Time-to-pandemic'' at the prediction (MLE, using cases instead of ``$n$''). \label{fig8}}
\end{figure}

\begin{figure}
%\resizebox{5in}{5in}{figure}
\rotatebox{0}{\resizebox{5in}{5in}{\includegraphics{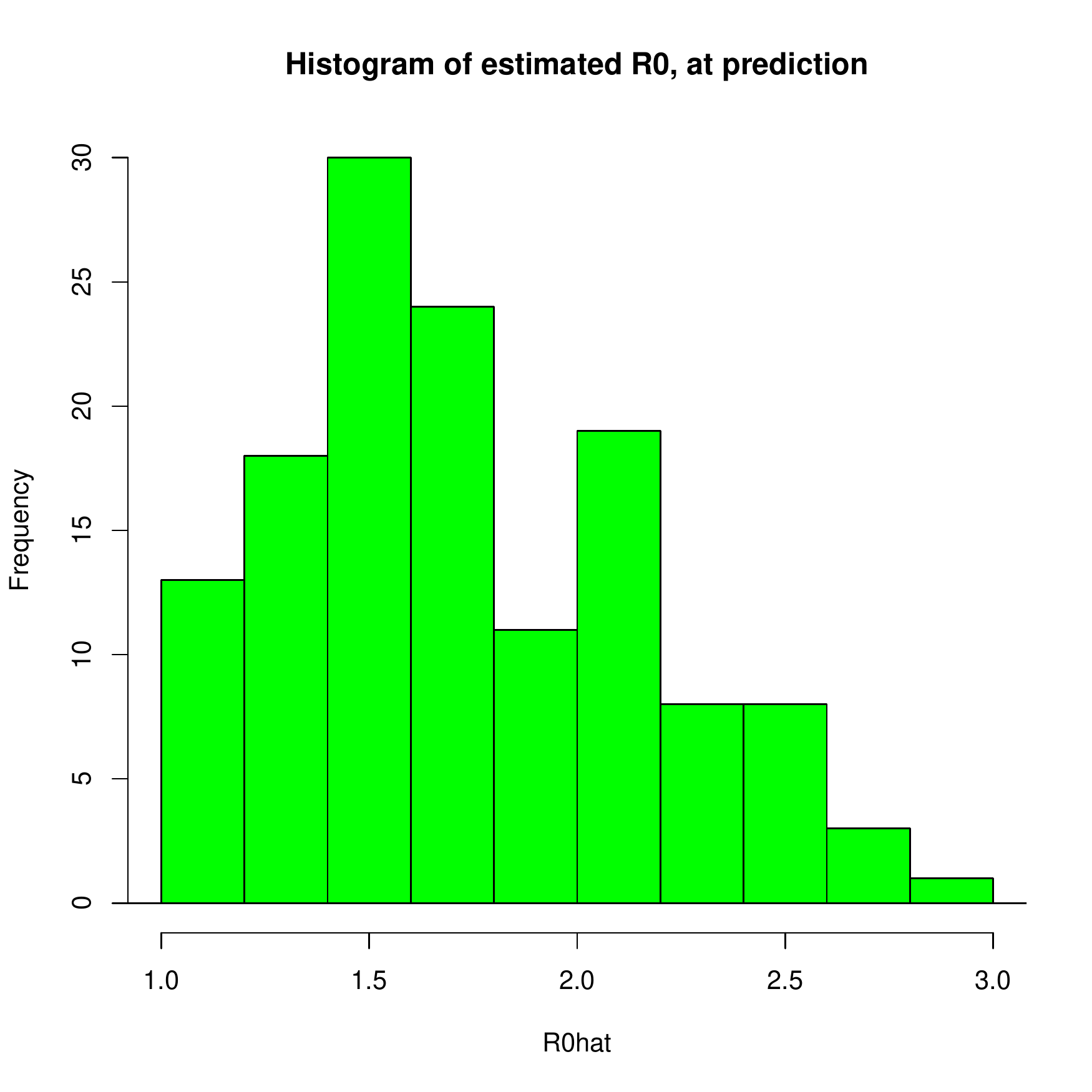}}}
\caption{Estimated R0's at the prediction. \label{fig9}}
\end{figure}

Figure \ref{fig6} shows a histogram of  
$n$\footnote{The bump at $n = 0$ is Fig. \ref{fig6} is due to starting the analysis of 
the $\{n_i\}$'s just after when total cases exceeded 200, to avoid lots of unfittable instances. Frequently, the first analysis generated a significant $\ro > 1$, halting the search. 
Since most cases are low-spreaders who generate no secondaries, the last $n$ was often zero. Incident cases can be zero, too, 
because the disease in humans had gone temporarily extinct.
The period explored was between 200 and 400 cases.} 
and Fig. \ref{fig7} the times-to-pandemic,\footnote{The sporadic period was three days.} made using the MLE-with-infectiousness-periods method,
for runs with p-value less than 0.1.
Note that many of the latter are large, even years. A somewhat more conservative prediction results by replacing ``$n$'' in equation (\ref{time_eqn}) by the number of cases existing
at the prediction point (counting secondaries created by the superspreader), producing Figure \ref{fig8}.
Figure \ref{fig9} shows the estimated $\ro$'s at the prediction.\footnote{Fig. \ref{fig9} seems to show a right-end tail, whose origin in this complicated context eludes me.}

Two other issues will produce more bias, in the familiar direction. 
As pointed out in paper I, a long ``kindling period'' when sporadic cases generate infection clusters that disappear will surely
lead anyone unaware of extra-Poisson variation and extinction probabilities to imitate the managers of the Space Shuttle program before the Challenger catastrophe
(who, after each launch without a disaster, were reassured about their risk judgement). Finally, secondary cases missing at random will produce bias 
towards underestimating $\ro$. 

To summarize: I have presented a small study, using simulations from mathematical models and conventional statistical ideas, on the topic of whether we should be able to
predict the occurrence of a SuperSpreader epidemic of the SARS/MERS type given contact data. 
The main conclusion is that care must be taken that mathematical modeling and statistical methodology do not add
to a false sense of security which will arise anyway from the long ``kindling period'' and certain
other, non-mathematical biases we all share. 
The methods I propose in this paper can overcome the statistical bias problems, granted adequate data and some modeling assumptions.\footnote{Many of these 
assumptions can be dropped without causing problems for the statistical methods. 
For instance, the two-level model was chosen for convenience, and can be easily
replaced by, e.g., the 
power-law distribution from paper II, with appropriate changes in the likelihood when using the ML method.} 
That a pandemic can often be predicted long before it begins is perhaps not so surprising; think of predicting when somebody will win in a repeated drawing in a lottery,
given facts such as how many people are buying tickets and the number of winning balls in the urn. Fitting a model is a means
of finding those facts; then mathematics (meaning today, computers) allows us to make the prediction. 

Re those ``other'' sources of error,
consider what psychologists call ``explanatory bias'' (but which might better be called
``explanatory hubris''). I refer to the universal tendency to imagine that one has found the 
explanation for an unusual or shocking event, and hence to believe that it was a one-time
occurrence or that it can easily be prevented in the future. 
({\em Viz}. that use of the aspirator in the ER that produced 140+ SARS cases in the Prince of Wales Hospital
in Hong Kong in 2003, or the putative wave of camel births in Saudi Arabia in the Spring of 2014, which somehow didn't
cause a bump in MERS cases a year earlier.) No one, including myself, desires to use stochastic modeling when a deterministic explanation can be found. 
But sometimes we can do better by incorporating randomness into our theory as a preventive for hubris.
Finally, there is the bureaucratic tendency to err on the side of not creating panic or generating reports that might displease the boss. 
Governments in the past have gone as far down this road as to deny
the occurrence of earthquakes, nuclear meltdowns, or disease outbreaks. 

In regards to MERS in 2014, as I write governments have behaved more responsibly in reporting cases to WHO than
was true for SARS in 2003, which is a step in the right direction. Epidemiologists cannot be expected to give sound advice to governments if they have to work from inadequate or
distorted data. MERS poses a potentially a greater risk to the world than did SARS, which went extinct after the zoonotic source, 
an exotic wild cat sold in live-animal markets in China, was eliminated.\footnote{As we have seen, extinction is a possibility that SS models, even with $\ro > 1$, permit.} 
If indeed the reservoir for MERS is camels, which are plentiful in the Middle East and Northern Africa,
re-introductions into the human population may continue for a long time. And the case mortality is frightful.

We modelers should not add to the false-sense of security problem, especially when thousands or millions of lives are at risk.  The results of this paper demonstrate that, with some 
attention to statistical issues,
modeling can make a contribution to preventing a global MERS epidemic.

 \section*{Appendix: statistical methods }

The model and fixed parameters (mean and variance of non-infectious and infectious periods) were described in previous papers. 
I modified the simulation routine to record the case index of primaries
that infected each secondary, then back-tracked to produce the $\{n_i\}$. If it were not for censoring, 
the statistical problem would be straighforward:
given an i.i.d. sample of nonnegative integers drawn from
a distribution:

\def\Pth{P_{\theta}}
\bar
\no \Pth\lsb\,n = k\,\rsb &\=& p\,\lrb{R_l^k\over k!}\rrb\,E\lrb\phi^k\,e^{-R_l\,\phi}\rrb +
(1-p)\,\lrb{R_h^k\over k!}\rrb\,E\lrb\phi^k\,e^{-R_h\,\phi}\rrb;\\
\phi &\=& \exp\lrb\,\sigma\,Z - {1\over2}\,\sigma^2\rrb,\label{dist_eqns}
\ear

\ni estimate the model parameters, denoted by letter $\theta$.  $Z$ is a standard normal random variable and $\sigma$ (fixed) 
is chosen to give the required variance (i.e., the true infection period for each case equals the mean infection period times $\phi$). $E$ denotes expectation over $Z$.
Here $\theta$ stands for either $(p,R_l,R_h)$ or, equivalently, $(R_l,R_h,R_0)$ where 

\be
R_0 \= p\,R_l + (1-p)\,R_h.
\ee

 I first tried the following, rather crude, non-parametric approach: I estimated the three parameters by

\def\phat{\hat{p}}
\def\Rhatl{\hat{R}_l}
\def\Rhath{\hat{R}_h}
\def\Rhato{\hat{R}_0}
\bar
\no N_l &\=& \sum\,1\lsb n_i <= 5\rsb;\\
\no \phat &\=& N_l/N;\\
\no \Rhatl &\=&\lrb 1/N_l\rrb\, \sum\,1\lsb n_i <= 5\rsb\,n_i;\\
\no N_h &\=& \sum\,1\lsb n_i > 5\rsb;\\
\no \Rhath &\=&\lrb 1/N_h\rrb\, \sum\,1\lsb n_i > 5\rsb\,n_i;\\
\Rhato &\=& \phat\,\Rhatl + (1-\phat)\,\Rhath.
\ear

\ni This simple scheme had too much bias, as seen in Fig. 1. 

So I next tried a \ml\ improvement. Defining the likelihood as usual:

\be
L = \prod_{i=1}^N\,\Pth\lrb n_i\rrb,
\ee

\ni and the score and Hessian by:

\bar
\no S_j &\=& {\partial \log(L)\over \partial \theta_j};\\
 H_{j,k} &\=& {\partial^2 \log(L)\over \partial \theta_j\,\partial \theta_k};\\
\ear

\ni I used the non-parametric estimates (assuming $N_l$ and $N_h$ were at least five) as starting points for Newton-Raphson iterations:

\be
\theta_j^{'} \= \theta_j \- \sum_k\,\lsb H^{-1} \rsb_{j,k}\,S_k.
\ee

\ni These iterations either failed immediately, because $\det(H)$ had the wrong sign (positive, when it should be negative), 
indicating lack of convexity of the likelihood surface,
or (in most runs) reduced the magnitude of the score by a factor of $10^{-4}$. 
(The expectation over $Z$ in equations (\ref{dist_eqns}) was performed numerically, using 200 evaluation points.)
Starting at, e.g., random choices of the $\theta_k$'s, usually failed by the determinant-sign, indicating that the likelihood is rarely globally convex for this problem,
raising lack-of-identifiability issues.

As remarked in the text, there was still considerable bias in the ML-improved estimates, so I tried a log transformation, redefining:

\def\ltr{\hbox{[logtrans]}} 
\def\Rbarl{\overline{R}_l}
\def\Rbarh{\overline{R}_h}
\bar
\no \Rbarl\ltr &\=&\lrb 1/N_l\rrb\, \sum\,1\lsb n_i <= 5\rsb\,\log(1+n_i);\\
\no \Rbarh\ltr &\=&\lrb 1/N_h\rrb\, \sum\,1\lsb n_i > 5\rsb\,\log(1+n_i);\\
\ear
 
\ni and then fitting the model by using the previous estimate for $\phat$ but solving the equations:

\bar
\no \sum_k \lpar\, \lrb{R_l^k\over k!}\rrb\,E\lrb\phi^k\,e^{-R_l\,\phi}\rrb\,\log(1+k)\,\rpar &\=& \Rbarl\ltr;\\
\no \sum_k \lpar\, \lrb{R_h^k\over k!}\rrb\,E\lrb\phi^k\,e^{-R_h\,\phi}\rrb\,\log(1+k)\,\rpar &\=& \Rbarh\ltr,
\ear

\ni for the $\hat{R}$'s by one-dimensional line searches. (The sums over $k$ can be truncated by careful treatment of the summand exponent, using Stirling.) 
Defining $\Rhato$ as usual gave Fig. \ref{fig3}.

With the amplified data $\{n_i,\tau_i\}$, the likelihood becomes simpler:

\be
L = \prod_{i=1}^N\,\Pth\lrb n_i,\tau_i\rrb,
\ee

\def\rhoth{\lsb \rho_h\,t \rsb}
\def\rhotl{\lsb \rho_l\,t \rsb}
\def\Pth{P_{\theta}}
\bar
\no \Pth\lsb\,n = k;\,\tau = t\,\rsb &\=& p\,\lrb{\rhotl^k\over k!}\rrb\,e^{-\,\rhotl} + \\
\no && (1 - p)\,\lrb{\rhoth^k\over k!}\rrb\,e^{-\,\rhoth};\\
\no \rho_h &\=& {R_h\over \hbox{Inf}};\\ 
\rho_l &\=& {R_l\over \hbox{Inf}};
\ear

\ni where Inf stands for mean infectious period. Again using the nonparametric estimates as starting points, the Hessian determinant had the right sign, 
Newton-Raphson iterates converged quickly and reduced the score to zero.

Re confidence, I first tried leaving out one of the $\{n_i,\tau_i\}$, but the resulting confidence intervals were absurdly narrow. I then tried resampling,
but now they were too wide. Moreover, I don't have a grasp of what such confidence statements are supposed to mean; what we really want to know is
how likely was the observation that $\ro > 1$, if in truth it wasn't.
So I retreated to making p-values by simulation, as follows. Given the estimated lower-type probability, $p$, I randomly chose $\rl$ and $\rh$ so that $p\,\rl + (1-p)\,\rh = 1$,
starting with $\rh$ uniform on $[3,0.95/(1-p)]$, 1,000 times. For each choice, I simulated a trajectory and 
refit the model to the mock data, generating a new $\ro$; if this $\ro$ was equal to or greater than
the original estimate, it counted as rejecting the null hypothesis when it was true. Then the p-value was set to: (count)/1,000. 
The method had power, with a type-I error rate of 0.1 and 1,000 cases, to reject the null when it was false (namely for the illustrative model, with $\ro = 1.4$), of 0.77.   

\vskip0.1in
C code is available from the author on request.
\vskip0.1in

For you statisticians who might be interested: I did consider several other ideas/improvements. 
In the book \cite{book08} I advertised the kNN method of model fitting, which arose in computer science back in the 1950s for problems in 
pattern recognition. This ``generic'' method was used in paper I for fitting SS epidemic models to some MERS cluster data, and in paper II to case data, 
but as it does not directly aim at estimating $\ro$ it may not be convincing
to many people.
And, as kNN reduces to \ml\ if you can simulate fast enough, it should suffer from the bias problem I demonstrated in Fig. \ref{fig1}. On that issue, there is
a large literature on reducing bias in ML estimation. The idea is to modify the likelihood, perhaps by subtracting something which will cancel the bias.
However, all the suggestions I found in the literature (which are typically tested on trivial, exactly-solvable, examples) involve taking three or more derivatives of the loglikelihood.
My enthusiasm flagged at that point. 
Re identifiability for the $n_i$-only data, perhaps the power-law model from paper II would be an improvement. 
Or is there some simpler model with plausible biological interpretation but also with globally 
identifiable parameters, granted sufficient data?
(However, it is a mortal sin to abandon realism and mutilate a model simply in order to gain identifiability of parameters.)
I leave these problems to those interested, but remark that, noting the extreme skewness and outliers
in distributions generated by SS epidemics, any such methods should be thoroughly tested by simulation from an appropriate model.

\end{document}